\documentclass[lefteqn,showpacs,preprint]{revtex4}
\usepackage{amsmath}
\usepackage[usenames]{color}

\begin{document}
\draft
\title{The mechanism of the polarization dependence of the optical transmission in subwavelength metal hole arrays }
\author{Qian Zhao$^1$, Chao Li$^1$, Yun-Song Zhou$^{1,\dag}$, Huai-Yu Wang$^{2,\ddag}$}
\date{\today }
\draft

\address{$^1$ Center of Theoretical Physics, Department of Physics, Capital Normal University. Beijing
100048, China}

\address{$^2$ Department of Physics, Tsinghua
University, Beijing 100084, China}

\begin{abstract}
\textbf{We investigate the mechanism of extraordinary optical
transmission in subwavelength metal hole arrays. Experimental
results for the arrays consisting of square or rectangle holes are
well explained about the dependence of transmission strength on the
polarization direction of the incident light. This polarization
dependence occurs in each single-hole. For a hole array, there is in
addition an interplay between the adjacent holes which is caused by
the transverse magnetic field of surface plasmon polariton on the
metal film surfaces. Based on the detailed study of a single-hole
and two-hole structures, a simple method to calculate the total
tranmissivity of hole arrays is proposed.}
\end{abstract}


\maketitle

\newpage
\textbf{1. Introduction}

The extraordinary optical transmission (EOT) [1,2] in a
subwavelength metal hole array is an interesting topic[3-12] because
its mechanism is still in exploring and it shows abundant features.
One of the features is that the transmissivity may depend on the
polarization direction of the incident light. Disclosing clearly the
reason behind the dependence is helpful to adjusting the EOT
strength, as well as to applying the EOT in optical devices. Lots of
experiments have been done for light in visible [3-7,13-17],
infrared [8-10], and terahertz [11,12,18,19] regions to observe the
dependence of EOT on polarization of the incident light. We here
sort both the arrays and holes into three kinds, respectively, as
summarized in Table 1. It is seen from Table 1 that among nine
structures, five have been fabricated to observe the dependence of
the EOT on the light polarization. In the second column, a square
lattice consisting of rectangle holes shows polarization dependence
while that consisting of square holes does not. In order to disclose
the reason behind the discrepancy, a theoretical investigation is
desirable.
\[
\]
Table 1. The experimental results that whether EOT is dependent on
the polarization of the incident light or not. The names in
parentheses are used in Sec.5.
\begin{tabular}{l|l|l|l}\hline
{ }&{\textbf{Square lattice}}&{\textbf{Rectangle
lattice}}&{\textbf{Single hole}}\\\hline \textbf{Square hole} &
{Independent[11,16] (S-S array)}
 &Unreported
(S-R array)& Unreported
\\\hline
\textbf{Rectangle hole} & Dependent[11,15-19] (R-S array) &
Unreported (R-R array) & Dependent[14,15]\\\hline \textbf{Circle
hole} & Independent[3-5,13] & Dependent[5-7,11,13] &
Unreported\\\hline
\end{tabular}

\[
\]

A lot of theoretical works about EOT in hole arrays have been
reported[20], they are mainly focused on the mechanism or factors
that cause in or influence EOT in hole arrays. A few of them
investigate the polarization dependence of EOT in hole array or
single hole [21-25] . Garcia \emph{et al}. [21] carried out a
rigorous solution of Maxwell's equations so as to obtain the
transmission of circle holes perforated in a thin perfect-conductor
screen for s and p polarization. Gordon \emph{et al}[22] explained
the polarization dependence in array of elliptic holes in terms of
the interaction between SPP and the periodic lattice grating.
Notwithstanding the approaches done, the systematical investigation
about the polarization dependence is still desirable.

In our opinion, when considering the EOT in an array consisting of
holes (or slits), there are basically the single-hole (slit) effect
and the inter-hole (slit) effect [26]. The former reflects the
transmissivity behavior of the light going through single
subwavelength holes, and the latter means the possible modulation of
transmissivity arising from the influence by neighboring holes.
Supposing that the polarization dependence does exist, then one
should know if the dependence is caused by the single-hole or
inter-hole effect or both.

In this paper we investigate the mechanism of the polarization
dependence of the EOT in a hole array consisting of square or
rectangle holes. Based on our simulation results by
finite-difference time-domain (FDTD) method [27], we reveal the
mechanism of the polarization dependence and explain the
experimental results. Furthermore, we find that it is possible to
get a simple way to calculate the transmissivity of the hole array,
which may avoid the burdensome simulation work in hole arrays.

This
paper is arranged as follows. In section 2 the hole-array model is
established. Before studying the EOT of the array, we study in
detail the EOT of a single-hole and double-hole structures in
sections 3 and 4, respectively, so as to clearly show the
single-hole and double-hole effects. Then the EOT in a hole array is
researched in section 5. In doing so, a simple method is proposed to
calculate the transmissivity of the hole array. In section 6 the
simple method proposed in Sec. 5 is applied to some arrays. It
appears that the application is satisfactory. Section 7 gives our
summary.
\\
\\
\textbf{2. The array model}

Our model is sketched in Fig. 1. Rectangle holes drilled on a metal
film form a two-dimensional lattice, consisting of rectangle cells,
in $xy$ plane. The lengths of two sides of each hole are $a$ and
$b$, and the lattice constants are $A$ and $B$, respectively. A
linearly polarized TM wave illuminates this structure along
$z$-direction. Before impinging the structure, the magnetic and
electric components are $H_{0}$ and $E_{0}$, respectively. In
simulation we always use
\begin{equation}
 E_{0}=H_{0}=0.31. 
\end{equation}

The angle between $y$ axis and $E_{0}$ direction is $\theta$.
Hereafter the light is termed as '' $\theta$-polarized''. From Fig.
1 the $x$ and $y$ components of electric and magnetic fields are
\begin{subequations}
\begin{eqnarray}
\left.\begin{array}{ll}
 E_{0x}=E_{0}\sin\theta \\
 E_{0y}=E_{0}\cos\theta
 \end{array}\right\}, \\
\left.\begin{array}{ll}
 H_{0x}=-H_{0}\cos\theta\\
 H_{0y}=H_{0}\sin\theta
 \end{array}\right\}.
\end{eqnarray}
\end{subequations}

The metal film with thickness $h=2\mu m$ is made of silver. The
incident wave length is $\lambda_{0}=0.6 \mu m$ . The dielectric
constant of silver vs. wavelength can be expressed as $\epsilon_
{Ag}=3.57-54.33\lambda_{0}^{2}+i(-0.083\lambda_{0}+0.921\lambda_{0}^{3})$[26,28].
Thus, as $\lambda_{0}=0.6 \mu m$,$\epsilon_ {Ag}=-15.989+i0.1491$.

When light goes through the array, the surface plasmon polariton
(SPP) will be excited in every hole and the metal surfaces. Since it
is a TM wave, after entering the holes, the electric field may have
a $z$-component, while the magnetic field does not. In each hole
there is a strong power, denoted as $P$. In simulation, this power
value $P$ is measured by a monitor M placed at the exit of the hole
labeled by "$0$". If the whole structure is removed, the power
measured by this monitor at the same place is denoted as
$P_{0}$[20,22].The transmissivity of this hole is defined as
$T=P/P_{0}$.

Our simulated result is the transmissivity $T$. Since $T$ is simply
linearly proportional to Poynting vector $S$, we will analyze the
construction of $S$ to explain the expression of $T$ obtained by
simulation.

Since the array consists of holes, the light behavior in any one
hole and the correlation between holes are essential in realizing
the light behavior when light going through the whole array.
Therefore, before studying the whole array, we explore the light
behavior when it goes through only one hole and a two-hole
structure.
\\
\\
\textbf{3. The single-hole structures}

Letting all the holes in the array except the one labeled by "$0$"
in Fig. 1 be closed, we set up a one-hole structure. We will study
the cases where the hole is a square and a rectangle, respectively.

Suppose that the amplitudes of the components of electromagnetic
field in the hole are $E^{hole}_{x}$, $E^{hole}_{y}$,
$H^{hole}_{x}$and $H^{hole}_{y}$, respectively. It is found from the
simulated results that these amplitudes as functions of angle
$\theta$ can be expressed by following way:
\begin{subequations}
\begin{eqnarray}\left.\begin{array}{ll}
 E^{hole}_{x}=E^{hole}_{0,x}\sin\theta\\
 E^{hole}_{y}=E^{hole}_{0,y}\cos\theta
 \end{array}\right\},\\
\left.\begin{array}{ll}
 H^{hole}_{x}=-H^{hole}_{0,x}\cos\theta\\
 H^{hole}_{y}=H^{hole}_{0,y}\sin\theta
 \end{array}\right\}.
\end{eqnarray}
 \end{subequations}
Equation (3) tells us that, when studying a single rectangle hole,
one merely needs to measure the field components at an arbitrary
polarization angle $\theta$ so as to get the
amplitudes$E^{hole}_{0,x}$, $E^{hole}_{0,y}$, $H^{hole}_{0,x}$and
$H^{hole}_{0,y}$in the hole by Eq.(3). Then the field components at
any other angle $\theta$ can be easily calculated in terms of
Eq.(3).

It is noticed that the angular dependence of the left hand side of
Eq. (3) is identical to that of Eq. (2). Based on this fact, we
introduce a concept of SPP polarization excitation ratios (PERs) of
each field component: the ratio of amplitude of field component in
the hole to that in vacuum. They are denoted by $\eta_{Ex}$,
$\eta_{Ey}$, $\eta_{Hx}$, and $\eta_{Hy}$ respectively, and
expressed as follows:
\begin{subequations}
\begin{eqnarray}\left.\begin{array}{ll}
\eta_{Ex}=E^{hole}_{x}/E_{0x}=E^{hole}_{0,x}/E_{0}\\
\eta_{Ey}=E^{hole}_{y}/E_{0y}=E^{hole}_{0,y}/E_{0}\end{array}\right\},\\
\left.\begin{array}{ll}
\eta_{Hx}=H^{hole}_{x}/H_{0x}=H^{hole}_{0,x}/H_{0}\\
\eta_{Hy}=H^{hole}_{y}/H_{0y}=H^{hole}_{0,y}/H_{0}\end{array}\right\}.
\end{eqnarray}
\end{subequations}
For example, $\eta_{Ex}$ is called the \emph{x}-component SPP
electric field PER. Since the SPP electromagnetic field PERs are
independent of angle, they are used to describe the basic property
of the hole.

As long as the field amplitudes in the hole are measured, the total
curve of the transmission power or transmissivity can be obtained.
To explain the transmissivity, one needs to calculate Poynting
vector $\textbf{S}=\textbf{E}\times\textbf{H}$.

In the hole, the averaged value in a time period of
\emph{z}-component of Poynting vector is

\begin{subequations}
\begin{eqnarray}
S^{hole}_{z}=(E^{hole}_{x}H^{hole}_{y}-E^{hole}_{y}H^{hole}_{x})/2,\\
S^{hole}_{z}=(E^{hole}_{0,x}H^{hole}_{0,y}\sin^{2}\theta+E^{hole}_{0,y}H^{hole}_{0,x}\cos^{2}\theta)/2.
\end{eqnarray}
\end{subequations}

Let us first investigate the case of a square hole. The parameters
are taken as $a=b=0.2\mu m$. The power measured in this hole is
denoted as $P^{hole}$ and the transmissivity of the hole is defined
as $T_{0}=P^{hole}/P_{0}$. The simulated transmissivity values are
plotted in Fig. 2 by crosses. Figure 2 shows that as $\theta$ angle
changes, $T_{0}$ varies between 19.31 and 19.38. Since the variation
scope is within calculation error, the transmissivity is regarded as
unchanged, i.e., it is independent of the polarization.

The calculated amplitude values of \emph{x} and \emph{y} components
of electromagnetic fields of SPP in the hole are presented in Fig. 2
by square and circle symbols. The results are easily fitted by
curves in terms of Eq. (3). Here, we have
\begin{subequations}
\begin{eqnarray}
E^{hole}_{0,x}=E^{hole}_{0,y}=1.88, \\
H^{hole}_{0,x}=H^{hole}_{0,y}=1.21.
\end{eqnarray}
\end{subequations}
Equation 6 exhibits two features.One is that the dimensions of the
field component amplitudes $E^{hole}_{0,x}$, $E^{hole}_{0,y}$,
$H^{hole}_{0,x}$and $H^{hole}_{0,y}$in the hole are larger than
those outside the hole $E_{0x}$, $E_{0y}$, $H_{0x}$and $H_{0y}$
respectively. This reflects the EOT character, i.e., the
transmissivity $T_{0}$ is greater than 1, as a subwavelength hole
should have. The other is that the electromagnetic fields in a
square hole behave as isotropic. This is an important feature of a
square hole, which has been discovered by experiments, as shown in
Table. 1. Later we will see that for a rectangular holes it is not
so.

For a square hole, substituting Eq. (6) into Eq. (5b), one obtains
\begin{equation}
S^{hole}_{z}=E^{hole}_{0,x}H^{hole}_{0,y}/2=E^{hole}_{0,y}H^{hole}_{0,x}/2,
\end{equation}
Here $S^{hole}_{z}$is independent of angle $\theta$, which is the
reason why the transmissivity $T_{0}$ is independent of angle
$\theta$ as shown in Fig.2.

Equation (6) leads to
\begin{eqnarray}
\left.\begin{array}{ll}
\eta_{Ex}=\eta_{Ey}\\
\eta_{Hx}=\eta_{Hy}
 \end{array}\right\}
\end{eqnarray}

In other words, along the two sides of the square hole, the SPP PERs
are equal. Equation (8) is the physical reason that the transmission
power is independent of the polarization direction in a square hole.

Equation (8) manifests the $\pi/2$ rotation symmetry of a square
hole. It is probably that if the $\pi/2$ rotation symmetry is
broken, Eq. (8) will not be valid. Consequently the transmission
power should change with the polarization angle.

Next we investigate the case of a rectangular hole. Experiments
showed the polarization dependence of transmission in rectangular
hole, see Table 1. Let us see our simulated results.

The simulation results of a rectangle hole are plotted in Fig. 3
where the two sides of the hole are $a=0.2\mu m$ and $b=0.1\mu m$,
respectively. The crosses in Fig. 3(a) show the transmissivity
$T_{0}$. It changes with the polarization angle. The symbols in Fig.
3(b) denote the field amplitudes in the hole, which can be fitted
with Eq. (3) but $E^{hole}_{0,x}=0.69\times10^{-2}$,
$E^{hole}_{0,y}=2.89$, $H^{hole}_{0,x}=1.21$ , and
$H^{hole}_{0,y}=0.53\times10^{-2}$. With these data we obtained:
$\eta_{Ex}=0.022$, $\eta_{Ey}=9.32$, $\eta_{Hx}=3.90$, and
$\eta_{Hy}=0.017$. That is to say,
\begin{eqnarray}
\left.\begin{array}{ll}
\eta_{Ex}\neq\eta_{Ey}\\
\eta_{Hx}\neq\eta_{Hy}
\end{array}\right\}
\end{eqnarray}
The SPP PERs in \emph{x} and \emph{y} directions for either electric
or magnetic field differs from each other. This feature is different
from that of a square hole. In each direction, when the electric
field is strong, then the magnetic field is weak, or vice versa.

The above discussions demonstrate that the SPP polarization
excitation ratios are the key roles to exhibit the properties of the
transmission with the polarization angle.

As has been mentioned above, one can choose an arbitrary
polarization angle $\theta$ to get the amplitudes $E^{hole}_{0,x}$,
$E^{hole}_{0,y}$, $H^{hole}_{0,x}$and $H^{hole}_{0,y}$. In this way,
one saves a lot of time and workload to avoid measuring the whole
angle region as shown by solid and open symbols in Figs. 2 and 3.

Since for a square hole, the SPP PERs in two axis directions are the
same, while for a rectangle one it is not, when changing the ratio
\emph{b/a}, the SPP PERs should vary. We keep $a=0.2\mu m$, and
change \emph{b} from 0.1 to $0.3\mu m$. The simulated SPP PERs are
displayed in Fig. 4 by symbols. The lines in Fig. 4 are just for
guiding eyes. When the ratio \emph{b/a} is small, say for $b=0.1\mu
m$, one of $\eta_{x}$ and $\eta_{y}$ is negligible compared to
another. It seems that in this case the field components
$E^{hole}_{x}$ and $H^{hole}_{y}$ are totally depressed. This
implies that, the SPP wave is mainly polarized with the magnetic
field along the longer side of the rectangle hole. In other word,
with respect to the polarization properties, the rectangle hole is
somehow equivalent to a slit when the ratio \emph{b/a} of the hole
is small. For convenience we refer to this kind of hole as
slit-hole. The character of a slit-hole is that the electric or
magnetic field component along one side direction is negligible
compared to the other side. According to our simulation results,
there exists a critical size for \emph{b} beyond which the PERs
$\eta_{x}$ and $\eta_{y}$are comparable to each other. For instance,
when $a=0.2\mu m$, it is found from Fig. 4 that the critical size
for $b$ is $b_{c}=0.18\mu m$. When $b>b_{c}$, the corresponding SPP
PERs rise suddenly so the character of a slit-hole disappears. In
other words, the SPP mode of electric field in \emph{x} direction
begins to excite. As the size of \emph{b} approaches that of $a$,
$\eta_{x}$ and $\eta_{y}$become closer. At $b=0.2\mu m$, the square
hole, the two solid lines meet at the value
$\eta_{Ex}=\eta_{Ey}=6.06$, and the two dashed lines meet at the
value $\eta_{Hx}=\eta_{Hy}=3.89$.
\\
\\
\textbf{4. The double-hole structures}

 Now we turn to investigate the inter-hole effect. For
this purpose we set up a two-hole structure by closing all the holes
in the array except the two labeled by "0" and "1", hereafter
referred to as 0-1 structure.

Before starting the investigation of the two-hole structure, let us
briefly retrospect the EOT of a double-slit structure[20]. The SPP
wave excited in one slit will interfere with that coming from the
other slit. The interference varies with the inter-slit distance
\emph{D}. This interference is the so-called inter-slit effect. As a
consequence, the total power passing through the structure
oscillates with the inter-slit distance \emph{D}. Each peak of the
power curve corresponds to the in phase interference between the two
slits.

A double-hole structure resembles a double-slit structure in that
there is an interference between SPPs excited in the two holes as
the SPP waves travel along the metal film surfaces, and the
interference varies with the inter-hole distance. Therefore, at
appropriate inter-hole distances, the interference will generate
largest transmission power. In simulation, we find that one of such
distances is $A=0.4\mu m$ when $a\times b=0.2\times 0.2\mu m^{2}$
and the polarization of the incident wave is $\theta=90^{o}$.

The transmissivity measured in this structure is denoted as
$T_{01}$. When fixing $A=0.4\mu m$, the variation of $T_{01}$ as a
function of the polarization angle $\theta$ is displayed by the
solid circles in Fig. 5(a). The data are well fitted with the
equations
\begin{equation}
T_{01}(\theta)=T_{0}+\Delta T_{01}(\theta),
\end{equation}
 where
\begin{equation}
\Delta T_{01}(\theta)=6.64\sin^{2}\theta,
\end{equation}
and $T_{0}= 19.36$. It should be noticed that $T_{0}$ is just the
transmissivity of single square hole structure, see Fig. 2. Equation
(10) clearly demonstrates that the total transmissivity comprise two
parts coming from the single hole and inter-hole effects,
respectively. The contribution from the inter-hole effect on the
transmissivity is expressed by Eq. (11). When $\theta=0$, the
inter-hole effect vanishes.

 Here we intend to disclose which factors the coefficient 6.64 of the interference term involves.
 To do so we analyze the behaviors of the SPPs traveling from one hole to the other along \emph{x} direction.
 That is to say, we should evaluate the Poynting vecter $S^{SPP}_{x}$ between the two holes.
 For this purpose we choose the mid-point between the centers of the two holes on the exit surface of
 the film as an observation point to measure $S^{SPP}_{x}$. Now we close the hole "1". Thus the system degenerates to a single-hole structure.
 In this case the SPPs at the observation point comes from the open hole. The observed $S^{SPP}_{x}$reads
 $S^{SPP}_{x}=E^{SPP}_{y}H^{SPP}_{z}-E^{SPP}_{z}H^{SPP}_{y}$. Our simulated results reveals that the
 absolute values of $E^{SPP}_{y}$ and $H^{SPP}_{z}$are negligible compared to $E^{SPP}_{z}$ and $H^{SPP}_{y}$.
 Thus we have $S^{SPP}_{x}=-E^{SPP}_{z}H^{SPP}_{y}$. Consequently, in the following, we merely take into account the contribution of
 $E^{SPP}_{z}$ and $H^{SPP}_{y}$, omitting those of $E^{SPP}_{y}$
and $H^{SPP}_{z}$.The simulated amplitudes of $E^{SPP}_{z}$ and
 $H^{SPP}_{y}$ at the observation point are displayed in Fig. 5(b) by
solid and open circles, respectively. The solid curves in Fig.5(b)
are plotted with
\begin{eqnarray}\left.\begin{array}{ll}
E^{SPP}_{z}=E^{SPP}_{0z}\sin\theta \\
H^{SPP}_{y}=H^{SPP}_{0y}\sin\theta
\end{array}\right\}
\end{eqnarray}

where $E^{SPP}_{0z}=1.12$ and $H^{SPP}_{0y}=0.82$. Comparing Eqs.
(12) and (3), we find that the ratio
$H^{hole}_{y}/H^{SPP}_{y}=H^{hole}_{0y}/H^{SPP}_{0y}$ is independent
of angle $\theta$, so we define this ratio as $\gamma_{01}$:

\begin{eqnarray}
\gamma_{01}=\left\{
\begin{array}{ll}
H^{SPP}_{y}/H^{hole}_{y}=H^{SPP}_{0y}/H^{hole}_{0y},& when
H^{hole}_{0y}\neq0
\\
 0 ,&when H^{hole}_{0y}=0
\end{array}\right.
\end{eqnarray}
This ratio is regarded as the conversion of the transverse magnetic
filed on surface excited by the magnetic field in the hole. It is
determined by the geometry and physical parameters of the
single-hole structure. The subscript "01" in $\gamma_{01}$ means the
SPP traveling from holes 0 to 1. In the present case, $\gamma_{01}=
0.82/1.21=0.68$. When $H^{hole}_{0y}=0$, the $H^{SPP}_{y}$ can not
be excited, so we define $\gamma_{01}=0$. In this case the
inter-hole effect vanishes.

Now we are ready to disclose which factors the coefficient of
$\sin^{2}\theta$ term in Eq. (11) comprises. Let both holes in the
structure open. The SPP wave from hole 0 will arrive at hole 1, it
will interfere with the SPP excited in hole 1. The same process will
also occur at hole 0. In this case the total magnetic field at the
exit of hole "0" includes two parts: One is
$H^{hole}_{y}$contributed from "0" itself as an isolated hole, and
the other is $H^{SPP}_{y}$ contributed from "1". Thus the
y-component of magnetic field becomes $H^{hole}_{y}+H^{SPP}_{y}$.
Consequently, the Poynting vector in \emph{z}-direction is
$$
S^{hole}_{01,z}=E^{hole}_{x}(H^{hole}_{y}+H^{SPP}_{y})-E^{hole}_{y}H^{hole}_{x}=S^{hole}_{z}+E^{hole}_{x}H^{SPP}_{y},\eqno(14)
$$

Here $S^{hole}_{z}$ is just Eq. (5a), the Poynting vector without
the contribution of $H^{SPP}_{y}$. Substituting Eqs. (3) , (12), and
(13) into Eqs. (14) we get
$$
S^{hole}_{01,z}=S^{hole}_{z}+\Delta S^{hole}_{01,z},\eqno(15a)
$$
where
$$
\Delta S^{hole}_{01,z}=2S^{hole}_{01,z}\gamma
_{01}\sin^{2}\theta,\eqno(15b)
$$
The Poynting vector of the 0-1 structure $S^{hole}_{01,z}$ comprises
two parts: a term of a single-hole structure and a term reflecting
inter-hole effect between the two holes. The latter in turn is
related to $S^{hole}_{01,z}$ its self. Therefore, although an angle
factor $\sin^{2}\theta$ is separated, the coefficient of
$\sin^{2}\theta$ in Eq. (15b) should generally still contain
functions of angle $\theta$. Equation (15) is linearly proportional
to Eq. (10). Therefore, we can reasonably rewrite Eq. (11) in the
following form:
$$
\Delta
T_{01}(\theta)=T_{01}(\theta)C_{01}(\theta)\sin^{2}\theta,\eqno(16)
$$
$C_{01}(\theta)$is the coupling coefficient that reveals the
strength of inter-hole effect generated by total transmission power.
Combining Eqs. (10) and (16) one achieves
$$
T_{01}(\theta)=T_{0}+T_{01}(\theta)C_{01}(\theta)\sin^{2}\theta,\eqno(17)
$$
We emphasize that Eq. (17) is applicable to the double-hole
structure consisting of identical rectangle holes.

In the case of square holes, we have, from Eq. (11),
$T_{01}(\theta)C_{01}(\theta)=6.64$. Thus the two expression of the
two factors $T_{01}(\theta)$ and $C_{01}(\theta)$ are easily solved.
$$
C_{01}(\theta)=6.64/(T_{0}+6.64\sin^{2}\theta),\eqno(18)
$$
and
$$
T_{01}(\theta)=T_{0}/[1-C_{01}(\theta)sin^{2}\theta].\eqno(19)
$$
Please note that since $T_{0}=19.36$, much larger than the term
$6.64\sin^{2}\theta$,$C_{01}(\theta)$ in Eq. (18) is approximately a
constant. Indeed, if we select $\theta=\pi/2$, then
$C_{01}(\theta)=0.255$, and the calculated $T_{01}(\theta)$ is
plotted in Fig. 5(a) by dashed line. Apparently, this is a quite
good approximation.

It is worthy to point out that the inter-hole effect involves the
contributions from SPP waves of both surfaces of the metal film.
Equations (10), (17), and (19) have included the contribution from
both surfaces.

When the nearest neighbor (nn) hole is farther and at another
azimuth angle, the interference between the two holes will vary. As
an example to demonstrate this, we choose the two holes labeled by
"0" and "5" in the array depicted in Fig 1, while other holes are
closed, to discuss the next nearest neighbor (nnn) inter-hole
effect. The array constants are $A=B=0.4\mu m$ and $a\times
b=0.2\times 0.2\mu m^{2}$. The simulated transmissivity as a
function of the polarization angle are plotted in Fig. 6 by solid
points. The data are well fitted by solid curve in terms of
following expressions:
$$
T_{05}=T_{0}+\Delta T_{05},\eqno(20)
$$
$$
\Delta
T_{05}=T_{05}C_{05}(\theta)\sin^{2}(\theta+45^{\circ}),\eqno(21)
$$
and
$$
C_{05}(\theta)=-2.55/[T_{0}-2.55\sin^{2}(\theta+45^{\circ})].\eqno(22)
$$
Compared to the 0-1 structure, the 0-5 structure shows differentia
in two ways. One is that the phase shift comes from the fact that
the hole "5" is located at azimuth angle $45^{ \circ}$.
Correspondingly, there is the same phase shift in $H^{SPP}_{y}$
compared to Eq. (12):
$H^{SPP}_{y}=H^{SPP}_{0y}\sin(\theta+45^{\circ})$. Therefore when
$\theta=135^{\circ}$, there will be no propagation of $H^{SPP}_{y}$
between the holes "0" and "5", i.e., the inter-hole effect vanishes
at this angle. Indeed, from Eq. (21) the interference term is zero
at this angle. The other is that the figure $-2.55$ in Eq. (22) is
in the place of 6.64 in Eq. (18).

The discussion about the interference in the two-hole structures
above only concerns the azimuth. Another factor affecting the
interference is the distance between the two holes. Let the distance
between the two holes be \emph{r}. Then the transmissivity
oscillates with \emph{r}. This oscillation is embodied in the value
of \emph{C}. Our simulation results show that for present square
lattice of $A=B=0.4\mu m$, as $r=A$,
$T_{01}(\theta)C_{01}(\theta)=6.64$, which just corresponds to the
interference in phase; and when $r=\sqrt{2}A$, this distance makes
the interference out of phase so that the transmission is
suppressed, thus $T_{05}C_{05}(\theta)=-2.55$ is  minus. From Eq.
(21) we see that as $\theta=45^{\circ}$ the inter-hole effect term
is maximum, so that the transmissivity curve in Fig. 6 shows a
valley.

We have mentioned that in the 0-1 structure, calculated
$T_{01}(\theta)$ using a constant $C(\theta)=0.255$ approximates the
exact results quite well, as shown in Fig. 5. Here we again set a
constant $C(\theta)=-0.152$ to compute $T_{05}(\theta)$ and the
results are plotted in Fig. 6 by dashed line, which is almost
identical to the solid line.
\\
\\
\textbf{5. The hole arrays }

To simulate the transmission of a hole array is quite difficult for
a very large memory size is needed. However, the discussion about
the two-hole structures in Sec. 4 prompts us that an array can be
regarded as a combination of two-hole structures. Here we propose a
simpler method to treat the hole array.

As an example, we first consider a three-hole structure consisting
of the open holes labeled by "1", "0" and "3" while other holes
being closed in the array depicted in Fig. 1, referred to as 1-0-3
structure. In such a structure, if the transmissivity of the hole
"0", denoted as $T_{103}$, is measured, one has to consider
inter-hole effects between hole "0" and its two neighbors . Thus a
reasonable expression should be
$$
T_{103}(\theta)=T_{0}+\Delta T_{01}+\Delta
T_{03}=T_{0}+T_{103}(\theta)C_{01}\sin^{2}\theta+T_{103}(\theta)C_{03}\sin^{2}(\theta+180^{\circ})
\eqno(23)
$$
Note that similar to the cases of 0-1 structure and 0-3 structures,
the coefficients of the two interference terms should include a
factor of the total transmissivity $T_{103}(\theta)$. Since the
holes "1" and "3" are symmetric with respect to the hole "0", we
have $C_{01}=C_{03}$. The difference of $T_{103}(\theta)$ between
results of the calculation with Eq. (23) and simulation by FDTD
method is about 1\%.

From the example of the 1-0-3 structure it is concluded that for
each additional nn hole, one merely simply add a term to embody the
inter-hole effect, although the coefficient should be proportional
to the total transmissivity of the hole "0". This conclusion can be
extended into the whole array.

Now let the all holes in the array open. We calculate the
transimissivity of hole "0". There are four nn and four nnn
neighbors around this hole, labeled by "1" to "8", respectively. The
influence of the holes farther than the nnn ones is merged into the
inter-hole effect between hole "0" and the eight neighbors, so that
it needs not to consider. Thus, the transimissivity of hole "0"
reads
$$
T_{array}=T_{0}+\sum^{8}_{i=1}\Delta T_{0i}.\eqno(24)
$$
The second term in Eq. (24) includes the contribution from its all 8
neighboring holes.

When the hole array composes a square lattices with $A=B$.

First we study the case where all holes in the lattice are square,
referred to as S-S array. The transmission in hole "0" is denoted as
$T_{S-S}$.

The 0-1 and 0-5 structures have been studied in detail in Sec. 4.
According to the conclusions of the two structures, we easily put
down the terms of inter-hole effects contributed from all the eight
neighboring holes as follows:
$$
\Delta T_{01}=T_{S-S}C_{01}\sin^{2}\theta,\eqno(25a)
$$
$$
\Delta
T_{02}=T_{S-S}C_{02}\sin^{2}(\theta-90^{\circ})=T_{S-S}C_{02}\cos^{2}\theta,\eqno(25b)
$$
$$
\Delta
T_{03}=T_{S-S}C_{03}\sin^{2}(\theta-180^{\circ})=T_{S-S}C_{03}\sin^{2}\theta,\eqno(25c)
$$
$$
\Delta
T_{04}=T_{S-S}C_{04}\sin^{2}(\theta-270^{\circ})=T_{S-S}C_{04}\cos^{2}\theta,\eqno(25d)
$$
$$
\Delta T_{05}=T_{S-S}C_{05}\sin^{2}(\theta+45^{\circ}),\eqno(26a)
$$
$$
\Delta T_{06}=T_{S-S}C_{06}\sin^{2}(\theta-45^{\circ}),\eqno(26b)
$$
$$
\Delta T_{07}=T_{S-S}C_{07}\sin^{2}(\theta-135^{\circ}),\eqno(26c)
$$
$$
\Delta T_{08}=T_{S-S}C_{08}\sin^{2}(\theta-225^{\circ}).\eqno(26d)
$$
Here the azimuth of each neighboring hole is taken into account.
Since the four holes "1" to "4" have the same distance away from
"0", and the other four on the vertexes do so too, one naturally
gets:
$$
C_{01}=C_{02}=C_{03}=C_{04},\eqno(27a)
$$
$$
C_{05}=C_{06}=C_{07}=C_{08}.\eqno(27b)
$$
Inserting Eqs. (25)-(31) into (24), we obtain
$$
T_{S-S}=T_{0}+2T_{S-S}(C_{01}+C_{05}) \eqno(28)
$$
Obviously, the transmission is independent of $\theta$ angle. This
explains the experimental result of S-S array listed in Table 1.

If the parameters of the holes and lattice are the same as those in
Sec. 4, we have $C_{01}+C_{05}=0.103$, $T_{0}=19.35$, thus
$T_{S-S}=24.37$.

Next we study the case where the holes are square and the lattice is
rectangular, referred to as S-R array. The transmission in hole "0"
is denoted as $T_{S-R}$. Since in this case $B\neq A$, we define an
angle $\alpha$:
$$
\alpha=\arctan(B/A).\eqno(29)
$$
The angular dependences of $\Delta T_{0i},i=1,2,3,4$ are the same as
those in Eq. (25). One merely need to replace $T_{S-S}$in Eq. (25)
by $T_{S-R}$ to get the expression of $\Delta
T_{0i},i=1,2,3,4$.However, since $B\neq A$, Eq. (27a) is not valid
any more. We have following relationship:
$$
C_{01}=C_{03}\neq C_{02}=C_{04}.\eqno(30)
$$
As for the neighbors "5" to "8", the angular dependence of
interference terms are written as
$$
\Delta T_{05}=T_{S-R}C_{05}\sin^{2}(\theta+\alpha),\eqno(31a)
$$
$$
\Delta T_{06}=T_{S-R}C_{06}\sin^{2}(\theta-\alpha),\eqno(31b)
$$
$$
\Delta
T_{07}=T_{S-R}C_{07}\sin^{2}[\theta-(180^{\circ}-\alpha)]=T_{S-R}C_{07}\sin^{2}(\theta+\alpha),\eqno(31c)
$$
$$
\Delta
T_{08}=T_{S-R}C_{08}\sin^{2}[\theta-(180^{\circ}+\alpha)]=T_{S-R}C_{08}\sin^{2}(\theta-\alpha).\eqno(31d)
$$
The distance between the hole "0" and any one of these four holes is
the same as others. Hence Eq. (27b) is still valid. Inserting Eqs.
(30) and (31) into $ T_{S-R}=T_{0}+\sum^{8}_{i=1}\Delta T_{0i}$, we
have
$$
T_{S-R}=T_{0}+2T_{S-R}\{[C_{01}\sin^{2}\theta+C_{02}\cos^{2}\theta]+C_{05}[\sin^{2}(\theta-\alpha)+\sin^{2}(\theta+\alpha)]\}.\eqno(32)
$$
It is seen that the transmissivity of the S-T array depends on the
polarization angle $\theta$.

Thirdly we discuss the case where the holes are rectangular,
$a\times b=0.2\times0.1\mu m^{2}$ and the lattice is square,
$A=B=0.4\mu m$, the so-called R-S array. Since the transmission in
each rectangular hole is dependent on $\theta$, as manifested in
Fig. 3(a), the single-hole effect is enough to cause the dependence
of transmissivity $T_{R-S}$on the polarization angle. Besides, the
inter-hole effect also influences $T_{R-S}$. Apparently, in this
case we again have relationship Eq. (30). In the case of $a\times
b=0.2\times0.1\mu m^{2}$, the hole is regarded as a slit-hole, as
having been mentioned in the last paragraph of Sec. 3 in explaining
Fig. 4. Considering Eqs. (15) and (13), for a slit-hole, we have
$\Delta
S^{hole}_{01,z}=2S^{hole}_{01,z}\gamma_{01}\sin^{2}\theta=0$, so
$\Delta
T_{01}(\theta)=T_{01}(\theta)C_{01}(\theta)\sin^{2}\theta=0$.
Consequently $C_{01}=0$. Following the method as above, the
expression of the transmission is obtained as follows:
$$
T_{R-S}=T_{0}(\theta)+2T_{R-S}[C_{02}(\theta)\cos^{2}\theta+2C_{05}(\theta)\sin^{2}45^{\circ}\cos^{2}\theta],
$$
By our simulation, it is approximately that
$C_{02}(\theta)=0.07/\cos^{2}\theta$,
$C_{05}(\theta)=0.12/\cos^{2}\theta$. Thus we get
$T_{R-S}=39.2\cos^{2}\theta$. The feature of $T_{R-S}$ curve is the
same as the experimental result [17]. This result confirms the
validity of our calculation method.

Finally, for the case of a rectangular lattice comprising rectangle
holes, R-R array, we can use the same method to discuss the
transmission $T_{R-R}$. But we do not put down the formula. A
qualitative conclusion is obvious. Since both the single holes and
lattice are rectangular, it is sure that $T_{R-R}$ depends on the
polarization angle.
\\
\\
\textbf{6. Applications}

 Up to now we have discussed the six cases
in Table 1. The two kinds of single holes are studied in detail in
Sec. 3 and the four kinds of arrays are investigated in Sec. 5. The
mechanism of the polarization dependence of the transmission of each
case is explicitly disclosed. A simple method is proposed to
evaluate the transmissivity of the arrays. The physical meaning of
this method is that it exhibits the transmission is mainly from two
parts: the single-hole and inter-hole effects. The obvious advantage
of this method is that it reduces the workload greatly compared to
the simulation of the whole array.

Among four kinds of arrays in Table 1, two, S-S and R-S arrays, have
been investigated experimentally, while the other two, S-R and R-R
arrays, have not. Here we employ our method to calculate the
transmissivity $T_{S-R}$ and $T_{R-R}$ of S-R and R-R arrays. The
numerical results are provided for someone to test.

For an S-R array, we chose $A=0.4\mu m$, $B=0.3\mu m$, and $a\times
b=0.2\times 0.2\mu m^{2}$. In this structure, $C_{01}=0.255$ is
unchanged, and $C_{02}$ and  $C_{05}$ have to be estimated. The
transmissivities of 0-2 and 0-5 structures are simulated and fitted
by $T_{02}=T_{0}+0.12T_{02}\cos^{2}\theta$ and
$T_{05}=T_{0}-0.08T_{05}\sin^{2}[\theta-\arctan(0.3/0.4)]$, so we
get $C_{02}=0.12$ and $C_{05}=-0.08$. Then the transmissivity is
expressed by
$$
T_{S-R}=19.36/\{1-2(0.255\sin^{2}\theta+0.12\sin^{2}\theta)+2\times0.08[\sin^{2}(\theta-36.9^{\circ})+\sin^{2}(\theta+36.9^{\circ})]\}
$$
The calculated curve is displayed in Fig. 7(a).

For an R-R array, we take $A=0.4\mu m$, $B=0.3\mu m$ and $ a=0.2\mu
m$, $b=0.1\mu m$. For such a slit-hole, $C_{01}=0$. In 0-2 and 0-5
structures, the simulation results are
$T_{02}=T_{0}(\theta)+6\cos^{2}\theta$ and
$T_{05}=T_{0}(\theta)+6.1\sin^{2}(36.9^{\circ})\cos^{2}\theta$
respectively. Here $T_{0}(\theta)$ is the single hole transmissivity
expressed in Fig. 3(a) with the formula
$T_{0}(\theta)=24.3\cos^{2}\theta$. When writing formulas in the
coupling form, $T_{02}=T_{0}(\theta)+C_{02}T_{02}cos^{2}\theta$ and
 $T_{05}=T_{0}(\theta)+C_{05}T_{05}sin^{2}(36.9^{\circ})\cos^{2}\theta$, we
obtain $C_{02}(\theta)=0.2/\cos^{2}\theta$ and
$C_{05}(\theta)=0.023/\cos^{2}\theta$. It is found that $C_{02}$
and$C_{05}$ vary with $\theta$ and cannot be regarded as constants
now. This arises from that $T_{0}(\theta)$ varies with $\theta$ in
this structure. Equation (27b) still holds in this lattice.
$C_{06}(\theta)=0.023/\cos^{2}\theta=C_{07}(\theta)=C_{08}(\theta)$.
Thus the coupling equation is
$T_{R-R}=T_{0}(\theta)+2T_{R-R}[C_{02}(\theta)\cos^{2}\theta+2C_{05}(\theta)\sin^{2}(36.9^{\circ})\cos^{2}\theta]$.
The calculated $T_{R-R}$ is plotted in Fig. 7(b). Comparing Fig.
7(a) and (b), we see that the variation scope of $T_{R-R}$ is larger
than that of $T_{S-R}$, since the R-R array has a stronger
anisotropy than S-R array. Especially, $T_{R-R}$ can be zero at
$\theta=90^{\circ}$.
\\
\\
\textbf{7. Summary}

We have investigated the polarization dependences of the
transmission in square and rectangular lattices consisting of
different subwavelength holes. The filed components and
transmissivities of single-hole and double-hole structures are
computed by use of FDTD simulation method. The behaviors of the
transmission are explored and the corresponding mechanisms are
disclosed. Our basic point of view is that the total transmissivity
of a hole array is determined by the two basic factors:  the
single-hole effect and inter-hole effects. Based on the results of
these structure, a compact method is suggested and applied to
investigate the hole arrays. Our conclusions are summarized as
follows. (1) The SPP PERs are key roles in single hole. In a square
hole the SPP PERs along the two sides of the hole are equal, which
leads to two consequences. One is that the SPP wave in the square
hole is along the polarization direction of the incident light, and
the other is that the amplitude of the SPP wave is in proportion to
that of the incident light at any polarization angle. By contrast,
in a rectangle hole, the SPP PERs is not isotropic, which results in
that the amplitude of SPP in a rectangle hole cannot reserve a fixed
proportion to that of incident light with different polarization
angle. Therefore the transmissivity depends on the polarization
angle. (2) The transverse magnetic filed of the SPP wave on the
metal film surface plays a key role in the inter-hole effect. (3)
The total transmissivity of the hole array can be expressed with the
single-hole transmissivity plus the terms reflecting inter-hole
effects between the hole and its nn and nnn neighbors. (4) The
conclusion (3) provides a simple method to calculate the
transmissivity of the hole arrays. By this method we calculated the
polarization dependence of the transmissivity for S-R and S-S arrays
that are not reported in literatures.

\vskip8pt \textbf{Acknowledgements}

\vskip5pt This work is supported by the 973 Program of China (Grant
No.2011CB301801) and the National Natural Science Foundation of
China (Grant No. 10874124), and Natural Science Foundation of
Beijing (No. 1102012).





\[
\]
\centerline{\bf Figure captions}

\begin{description}
\item{Fig.1} The sketch of a metal hole array consisting of subwavelength rectangle
holes.

\item{Fig.2} The simulated transmissivity $T_{0}$ and polarization dependence of amplitudes of the electric and magnetic fields in a single-square-hole structure. The fitting curves are calculated by Eq. (3).

\item{Fig.3} The polarization dependence of quantities for a rectangle hole. (a) The transmissivity. The fitting curve is $T_{0}(\theta)=24.3\cos^{2}\theta$.  (b) The amplitudes of the electric and magnetic fields. The fitting curves are calculated by Eq. (3).

\item{Fig.4} The SPP polarization excitation rations of a rectangle hole with $a=0.2\mu m$ and $b$ varies from  $0.1$ to $0.3\mu m$. The lines are just to guide eyes.

\item{Fig.5} (a) The polarization dependence of the transmissivity $T_{01}$ for a two-hole structure. The fitting solid curve is from Eqs. (10) and (11). The dashed curve is obtain by the approximation of $C_{01}=0.255$. (b) The amplitudes of the electric and magnetic fields along the metal surface yielded from a single hole. The fitting curves are from Eq. (12).

\item{Fig.6} The simulated transmissivity in a 0-5 structure. The dashed curve is plotted by the fitted formula in the approximation of $C_{05}=
-0.152$.

\item{Fig.7} The calculated transmissivity in (a) the S-R array and (b) The R-R array.

\end{description}


\end{document}